\documentclass[aps,superscriptaddress,showpacs,floatfix]{revtex4-1}

\usepackage{hyperref}
\usepackage{color}
\usepackage{graphicx}	
\graphicspath{%
  {./},%
}

\newcommand{\name}{\mbox{SONIC}\xspace}

\newcommand{\pp}{\mbox{p+p}\xspace}
\newcommand{\cc}{\mbox{C+C}\xspace}
\newcommand{\alal}{\mbox{Al+Al}\xspace}
\newcommand{\auau}{\mbox{Au+Au}\xspace}
\newcommand{\pbpb}{\mbox{Pb+Pb}\xspace}
\newcommand{\cucu}{\mbox{Cu+Cu}\xspace}

\usepackage{xspace}	

\begin{document}



\title{Particle spectra and HBT radii for simulated central nuclear collisions of \cc, \alal, \cucu, \auau, and \pbpb from $\sqrt{s}=62.4-2760$ GeV}


\author{M. Habich, J. L. Nagle, and P. Romatschke} 
\affiliation{University of Colorado Boulder}
\date{\today}

\begin{abstract}
We study the temperature profile, pion spectra, and HBT radii in central, symmetric, and boost-invariant nuclear collisions, using a super hybrid model for heavy-ion collisions (\name), combining pre-equilibrium flow with viscous hydrodynamics and late-stage hadronic rescatterings. In particular, we simulate Pb+Pb collisions at $\sqrt{s}=2.76$ TeV, \auau, \cucu, \alal, and \cc collisions at $\sqrt{s}=200$ GeV, and \auau and \cucu collisions at $\sqrt{s}=62.4$ GeV. We find that \name provides a good match to the pion spectra and HBT radii for all collision systems and energies, confirming earlier work that a combination of pre-equilibrium flow, viscosity, and QCD equation of state can resolve the so-called HBT puzzle. For reference, we also show \pp collisions at $\sqrt{s}=7$ TeV. We make tabulated data for the 2+1 dimensional temperature evolution of all systems publicly available for the use in future jet energy loss or similar studies.


\end{abstract}

\maketitle

\section{Introduction}

With the advent of gauge/gravity duality, it has become possible to effectively simulate far-from equilibrium thermalization in central (and smooth) nuclear collisions \cite{vanderSchee:2013pia}. Combining this pre-equilibrium dynamics with hydrodynamics \cite{Luzum:2008cw} and a late-stage hadronic cascade \cite{Novak:2013bqa}, one obtains a 'Super hybrid mOdel simulatioN for relativistic heavy-Ion Collisions' (\name for short) that effectively has only a limited number of parameters, namely those specifying the properties of the incoming nuclei, the speed of sound, and shear and bulk viscosities in the quark--gluon plasma. In this work, we use this model to study symmetric nuclear collisions of different nuclei (Pb, Au, Cu, Al, C) at collision energies ranging from $\sqrt{s}=62.4$ GeV to $\sqrt{s}=2.76$ TeV. We study the temperature evolution, pion spectra, and HBT radii for these different collision systems with the aim of both testing the model against experimental data where available and providing model predictions for the design of future experimental studies. In addition, we also show results for \name for central \pp collisions at $\sqrt{s}=7$ TeV energy, even though evidence for forming an equilibrated quark--gluon plasma in these systems is currently lacking. 

A fundamental question regarding the quark--gluon plasma is at what temperature and what scale a strong coupling description is most appropriate versus weak coupling.  Near-inviscid hydrodynamic modeling indicates strong coupling, though the exact sensitivity of final state hadrons to the temperature dependence of $\eta/s$ is currently under investigation.   There are experimental observables when compared to model calculations that are potential indicators of stronger coupling at temperatures near the transition point.   Inclusion of this strongest coupling near the transition
is proposed to help reconcile the full suite of jet quenching observables including the anisotropy in mid-central collisions
~\cite{Li:2014hja,Renk:2014nwa}, the larger than expected $v_{2}$ and $v_{3}$ of direct photons~\cite{vanHees:2014ida} and heavy quark observables~\cite{Adare:2013wca}.    An important motivation
for the sPHENIX upgrade~\cite{Aidala:2012nz} is to answer the question regarding the underlying nature of the
quark--gluon plasma near the point of strongest coupling.   A key question is whether high statistics data sets in \auau collisions at $\sqrt{s}=200$ GeV at RHIC and \pbpb collisions at $\sqrt{s}=2.76$ TeV at the LHC are substantially augmented by hard process observables at lower RHIC energies and with different nuclear geometries for emphasizing emission and parton quenching interactions
closer or further away from this transition temperature.  In this work, we explore the temperature evolution of different systems and provide access to the space--time snapshots for utilization in jet quenching, photon emission, and
heavy quark diffusion calculations.

\section{Methodology}
\begin{figure}[t]
  \centering
  \includegraphics[width=0.5\linewidth]{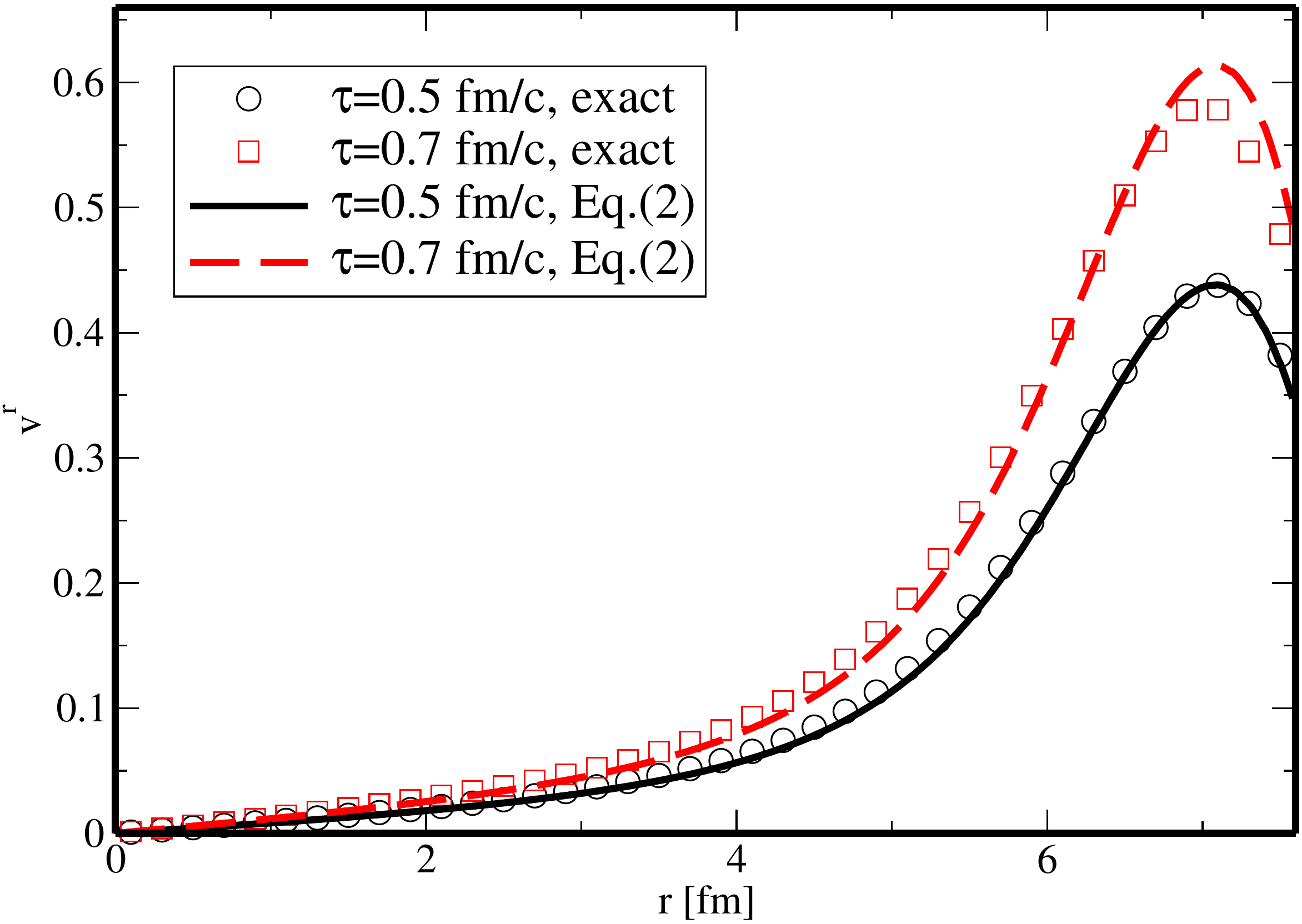}
  \caption{\label{fig:one} 
Comparison between the pre-equilibrium radial flow velocity obtained for \pbpb collisions at $\sqrt{s}=2.76$ TeV using a full numerical relativity simulation \cite{vanderSchee:2013pia} (``exact'') and the model equation (\ref{pre-flow}).
}\end{figure}

We model heavy-ion collisions by using a super hybrid model which we call \name which combines pre-equilibrium flow with hydrodynamics and a late-stage hadronic afterburner. Introducing the radius $r=\sqrt{x^2+y^2}$, the different nuclei are modeled by employing an overlap function 
\begin{equation}
\label{eq:overlap}
T_A(r)=\epsilon_0\int_{-\infty}^\infty dz \left[1+ e^{-(r^2+z^2-R)/a}\right]\,
\end{equation}
with $R,a$ the charge radius and skin depth parameters listed in Tab. \ref{tab:one}. $\epsilon_0$ is an overall normalization constant that controls the total final multiplicity. The pre-equilibrium flow has been calculated numerically assuming an infinite number of colors and infinite coupling for central (and smooth) Pb-Pb collisions at $\sqrt{s}=2.76$ TeV in Ref. \cite{vanderSchee:2013pia}. We re-analyzed the results from Ref.~\cite{vanderSchee:2013pia}, finding that after the system has thermalized, the velocity is consistent with the early-time analytic result derived in Ref. \cite{Romatschke:2013re} up to an overall factor of two (see Figure \ref{fig:one}). Therefore, in the following we will employ the pre-equilibrium radial flow velocity 
\begin{equation}
\label{pre-flow}
v^r(\tau,r)=-\frac{\tau}{3.0}\partial_r \ln T_A^2(r)\,, 
\end{equation}
where $\tau=\sqrt{t^2-z^2}$, see Fig.~\ref{fig:one}. Using Eq.~(\ref{pre-flow}) and an initial energy density profile given by (cf. Ref. \cite{Romatschke:2013re})
\begin{equation}
\label{pre-flowED}
\epsilon(\tau,r)=T_A^2(r)\,, 
\end{equation}
we start the hydrodynamic evolution at a time $\tau_{\rm sw}$. Note that we have chosen the energy density to scale as the overlap function squared because this scaling is known to give a simple description of the centrality dependence of multiplicity, cf. \cite{Kolb:2001qz}.
Following the observations in Ref.~\cite{vanderSchee:2013pia,Arnold:2014jva}, $\tau_{\rm sw}$ has to be large enough such that a local rest-frame can be defined as ($\tau_{\rm sw}\gtrsim 0.35$ fm/c) and before non-linear effects prohibit the use of Eq.~(\ref{pre-flow}) ($\tau_{\rm sw}\lesssim 0.6$ fm/c). 
\begin{table}[b]
\begin{tabular}{lcccc}
Isotope & $\sqrt{s}$ [GeV] & $R$ [fm] &$a$ [fm]  & $T_0(\tau=0.5 {\rm fm})$ [MeV] \\ \hline
p-1    & 7000 & --   & 0.4 & 390 \\
C-12 & 200 & 2.355 & 0.522 & 238 \\
Al-27 & 200 &3.061 & 0.519 & 287 \\
Cu-63 & 62.4 &4.163 & 0.606 & 300 \\
Cu-63 & 200 &4.163 & 0.606 & 327 \\
Au-197 & 62.4 &6.380 & 0.535 & 340\\
Au-197 & 200 &6.380 & 0.535 & 370\\
Pb-208 & 2760 &6.624 & 0.549 & 470
\end{tabular}
\caption{\label{tab:one} Model parameters for different collision systems \cite{DeVries1987495,DeJager1974479}. For all systems we use $T_S=170$ MeV, $\eta/s=0.08$, $\zeta/s=0.01$, and QCD equation of state at zero baryon density \cite{Laine:2006cp}. Parameters $R$ and $a$ correspond to Eq.~(\ref{eq:overlap}) except for p-1 where $a$ denotes the width of a Gaussian, i.e., $T_A(r)=\epsilon_0 \int dz e^{-r^2/2/a^2}$. }
\end{table}
Using the energy density from Eq.~(\ref{pre-flowED}), the flow profile from Eq.~(\ref{pre-flow}), and setting the initial shear and bulk stresses to zero, we can solve the subsequent system evolution using the relativistic viscous hydrodynamics solver VH2+1 \cite{Romatschke:2007mq,Luzum:2008cw}, version 1.7. The fluid shear viscosity over entropy ratio is set to $\eta/s=0.08$ and the bulk viscosity over entropy ratio is set to $\zeta/s=0.01$. The equation of state used is that from Ref.~\cite{Laine:2006cp} which is consistent with lattice QCD data \cite{Borsanyi:2013bia,Bazavov:2014pvz} at vanishing baryon density and matches a hadron resonance gas at low temperatures. 
We monitor the isothermal hypersurface defined by $T_S=170$ MeV throughout the system evolution until the last fluid cell has cooled below $T_S$. 

From the information about fluid temperature, velocity, and dissipative stress components we generate hadrons with masses up to 2.2 GeV and follow their rescattering dynamics using the hadron cascade code B3D \cite{Novak:2013bqa}. Details for the freeze-out procedure can be found in the original reference \cite{Novak:2013bqa}, but for completeness we mention that the particle spectra take into account deformations from shear and bulk stresses indepent of particle type as outlined in \cite{Pratt:2010jt} such that the full energy-momentum tensor is continuous accross the freeze-out hypersurface. We then generate 5000 B3D events for each hydro event. Once the particles have stopped interacting we collect information about the particle spectra and report the total charged multiplicity $\frac{dN_{\rm ch}}{dy}$, the mean pion transverse momentum $<p_T>$ and the pion HBT radii $R_{\rm out}$, $R_{\rm side}$, and  $R_{\rm long}$.

With the pre-equilibrium flow given by Eq.~(\ref{pre-flow}), and adjusting $\epsilon_0$ so that total multiplicity is constant, we find that the final particle $<p_T>$ and the extracted HBT radii are insensitive to the choice of $\tau_{sw}$, just as in the full gauge/gravity+hydro+cascade calculation (cf. Ref.~\cite{vanderSchee:2013pia}). Thus, $\tau_{sw}$ is not a relevant parameter of \name. This leaves a total of 6 relevant parameters for the system evolution: three numbers ($R$, $a$, $T_S$) and three functions (the temperature dependent ratios $\eta/s$, $\zeta/s$, and the equation of state). Note that $\epsilon_0$ is fixed by requiring the final charged multiplicity to match experimental data, wherever it is available
\cite{Aamodt:2010pp,Alver:2010ck,Aamodt:2010pb}, cf. Tab.~\ref{tab:two}.
For \cc and \alal collisions at $\sqrt{s}=200$ GeV, we employ the formula
\begin{equation}
\label{mult-form}
\frac{dN_{\rm ch}}{dy}=\left[\alpha(\sqrt{s}) N_{\rm coll}+\frac{1-\alpha(\sqrt{s})}{2}N_{\rm part}\right] \frac{dN_{pp}}{dy}\,,
\end{equation}
where $\frac{dN_{pp}}{dy}$ is the charged multiplicity for nucleon-nucleon collisions at a given collision energy $\sqrt{s}$, and $\alpha(\sqrt{s}=200\ \mathrm{GeV})=0.13$
(cf. Ref.~\cite{Back:2002uc}).

We note that a full 2+1 dimensional simulation of a single symmetric nuclear collision can be executed on a modern desktop in approximately one hour, which makes \name a viable tool to investigate collisions having granular initial conditions on an event-by-event basis in the future.

\begin{figure}[t]
  \centering
  \includegraphics[width=0.7\linewidth]{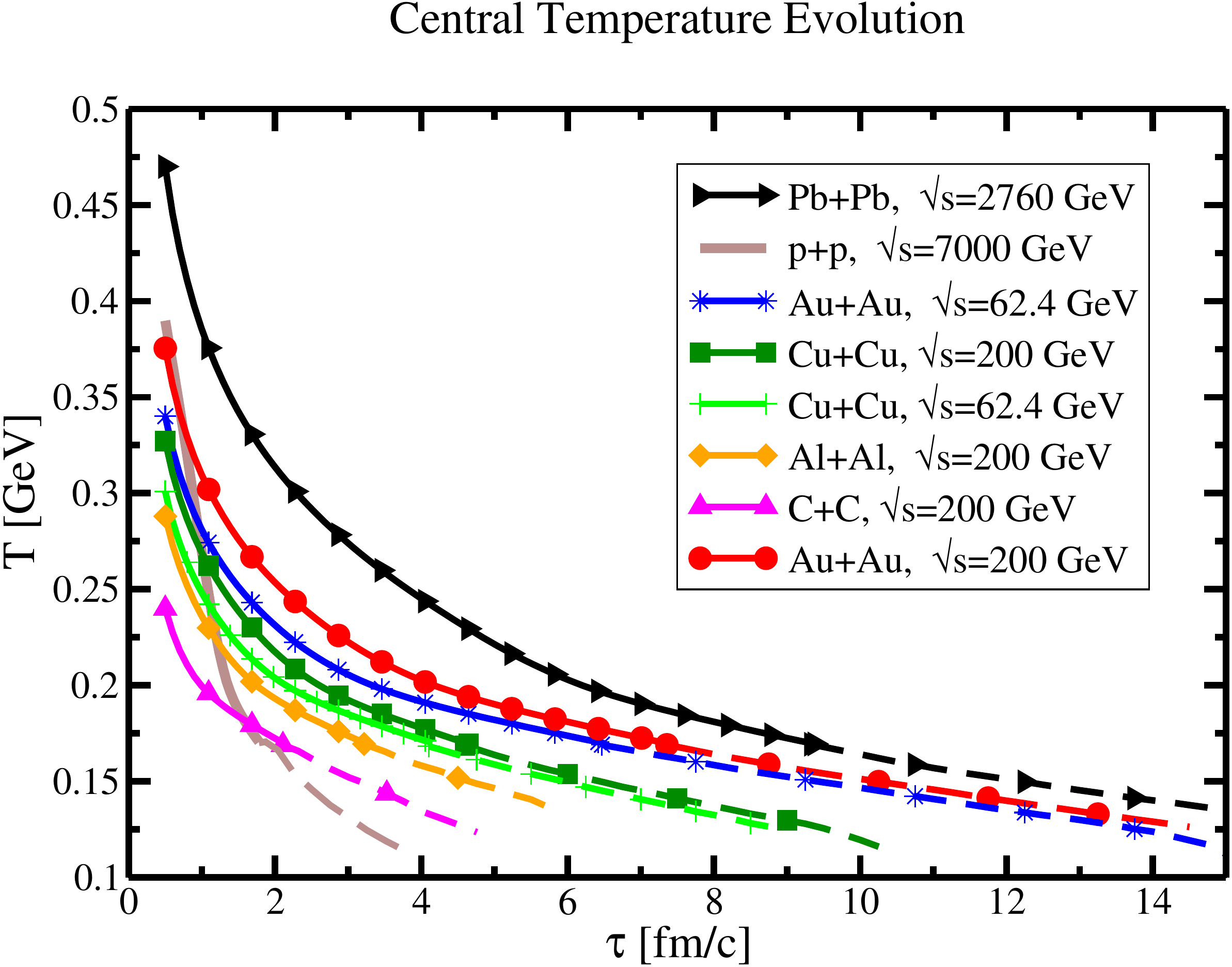}  
  \caption{\label{fig:two} 
Temperature evolution as a function of proper time at the center of the fireball ($r=0$) for different collision systems and different collision energies. Full lines denote evolution within hydrodynamics ($T>T_S$), dashed lines denote hadron gas regime ($T<T_S$). For reference, also $p+p$ collisions at $\sqrt{s}=7$ TeV are shown, even though this system may not equilibrate at all. The ``kink'' at $2$ fm/c in the temperature evolution in the $p+p$ system around $T=T_S$ is due to the center $r=0$ being cooler than the surrounding matter. }
\end{figure}
\begin{figure}
  \centering
  \includegraphics[width=0.7\linewidth]{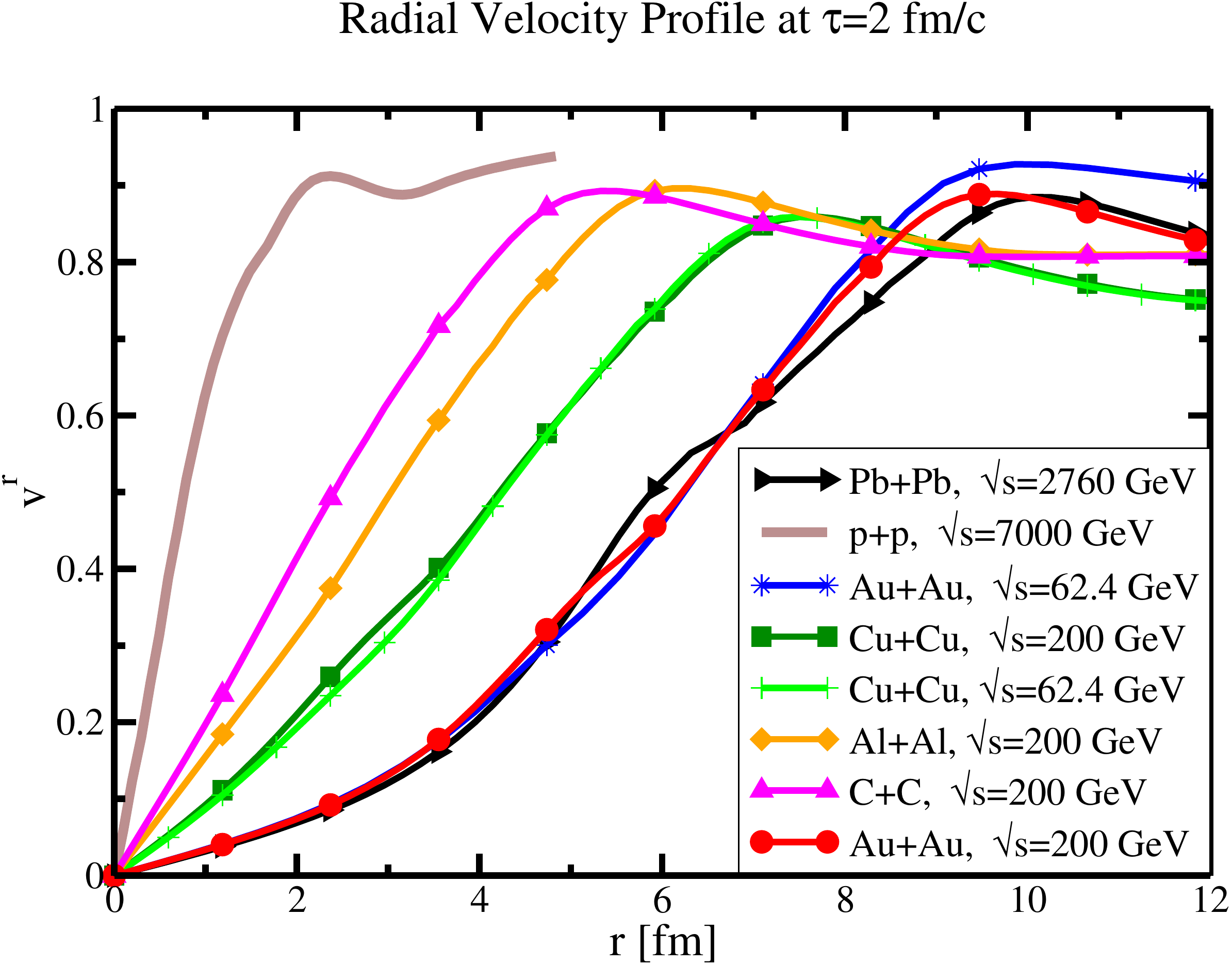}
  \caption{\label{fig:three} 
Velocity profile at $\tau=2$ fm/c for the different collision systems ($\tau=1.9$ fm/c for \pp). The velocity profiles for the \pbpb and \auau systems are similar because the systems have similar geometry and the final-observed larger radial flow at higher $\sqrt{s}$ is simply due to the longer lifetime of the \pbpb system.}
\end{figure}

\begin{table}
\begin{tabular}{lllllll}
Isotope & $\sqrt{s_\mathrm{NN}}$ [GeV] & $N_\mathrm{part}$ & $N_\mathrm{coll}$  & $\mathrm{d} N_\mathrm{ch}/\mathrm{d} \eta$ & $\langle p_\mathrm{T}\rangle$ [MeV] & Comments \\ \hline
p-1 & 7000 & - & - & 7 & 599 & for $p_T>0.15$ GeV\\
{\bf p-1 (exp.)} & {\bf 7000} & - & - & {\bf 6} & {\bf 622$\pm$21} & \cite{Aamodt:2010pp,Abelev:2013bla}, min.-bias\\
\hline
C-12 & 200  &17 & 19  & 21 & 396 & \\
Al-27 & 200 & 45 & 70 &  68 & 415 & \\
\hline
Cu-63 (th.)& 62.4 &111  &227  & 144 & 403 &  \\
{\bf Cu-63 (exp.)}& {\bf 62.4} & $106 \pm 3$  & $162 \pm 13$  &  {\bf 138}$\pm${\bf 10} & {\bf 379}$\pm${\bf 20}  & \cite{Alver:2010ck,Abelev:2009tp,Aggarwal:2010pj}, 0-10\% most central \\
\hline
Cu-63 (th.)& 200 &  113 & 227 & 193 & 421 &  \\
{\bf Cu-63 (exp.)}& {\bf 200} & $108 \pm 4$  & $189\pm14$  &  {\bf 198}$\pm${\bf 15} & {\bf 420}$\pm${\bf 20}  & \cite{Alver:2010ck,Abelev:2009tp,Aggarwal:2010pj}, 0-10\% most central\\
\hline
Au-197 (th.)& 62.4 & 375  & 1173  & 508 & 402  & \\
{\bf Au-197 (exp.)}& {\bf 62.4} & $356 \pm 11$  & -  &  {\bf 472 $\pm$41}  & {\bf 405$\pm$ 11.0}  &  \cite{Alver:2010ck,Abelev:2008ab}, 0-5\% most central\\ 
\hline
Au-197 (th.)& 200 & 378 & 1173 &677 & 424  & \\
{\bf Au-197 (exp.)}& {\bf 200} & $361 \pm 11$ &$1065 \pm 105$ &  {\bf 691$\pm$52} & {\bf 453$\pm$33}  & \cite{Adler:2003cb,Alver:2010ck}, 0-5\% most central \\
\hline
Pb-208 (th.)& 2760 & 399  & 1217  & 1635   & 503 & \\
{\bf Pb-208 (exp.)}& {\bf 2760} & $382 \pm 27$  & -  & {\bf 1584$\pm$80}   &  {\bf 517$\pm$19} & \cite{Aamodt:2010pb,Abelev:2013vea}, 0-5\% most central
\end{tabular}
\caption{
\label{tab:two}
Details for collision systems compared to experimental data. $<p_T>$ is for pion transverse momentum except for \pp collisions where report $<p_T>$ for $\pi,K,p$. We use $\frac{dN_{\rm ch}}{d y}=1.1 \frac{dN_{\rm ch}}{d\eta}$ to convert model multiplicity to pseudorapidity distribution.}
\end{table}

\section{Results}

Our results for the multiplicity and mean pion transverse momentum in the different systems are reported in Tab. \ref{tab:two} alongside with experimental results where available. Since the experimental multiplicity is used to fix one of the model parameters ($\epsilon_0$), only the pion $<p_T>$ is a non-trivial model output. Comparing experimental measurements of $<p_T>$ with model output from \name, we find that with model parameter choices $\eta/s=0.08$, $\zeta/s=0.01,T_S=170$ MeV, and a QCD equation of state, there is good agreement with experimental data for all collision systems at all collision energies.

 For reference, we also show \name runs for \pp collisions at $\sqrt{s}=7$ TeV collision energy, even though this system may not form an equilibrated state of matter (and thus \name would not be applicable in this case). Note that, nevertheless, the $<p_T>$ value for \pp collisions is not too far from the experimental value, which may just be a reflection of the fact that transverse flow is not an indicator of system equilibration (cf.~\cite{vanderSchee:2013pia}). 

It should be noted that our current implementation of SONIC does not properly reproduce the experimentally measured proton spectra because the number of protons and antiprotons is too high. The reason for this has been identified to be the missing implementation of baryon--antibaryon annihilation \cite{Pan:2012ne,Song:2013qma}. For this reason, we currently are unable to report physically viable results for baryons.

The time evolution for the temperature in the center of the fireball ($r=0$ fm) is reported in Fig.~\ref{fig:two}, where we distinguish between the evolution spent in the hydrodynamic phase ($T>T_S$) and the hadron gas phase at low temperature. Shown in Fig.~\ref{fig:three} is the radial velocity profile for the different collision systems at $\tau=2$ fm/c inside the hydrodynamic phase. Not surprisingly, larger systems tend to build up smaller radial flow and tend to live longer than smaller systems. However, possibly interesting features for temperature evolution between different systems may also be identified in Fig.~\ref{fig:two}. For instance, note that Fig.~\ref{fig:two} implies that the central temperature evolution in \auau collisions at $\sqrt{s}=62.4$ GeV starts out close to the \cucu $\sqrt{s}=200$ GeV results, but then eventually approaches the \auau $\sqrt{s}=200$ GeV curve. We are making the full two-dimensional space--time evolution profiles available for potential use in studies of jet energy loss, direct photon emission and heavy quark diffusion~\cite{data}.

\begin{figure}[t]
 \includegraphics[width=0.49\linewidth]{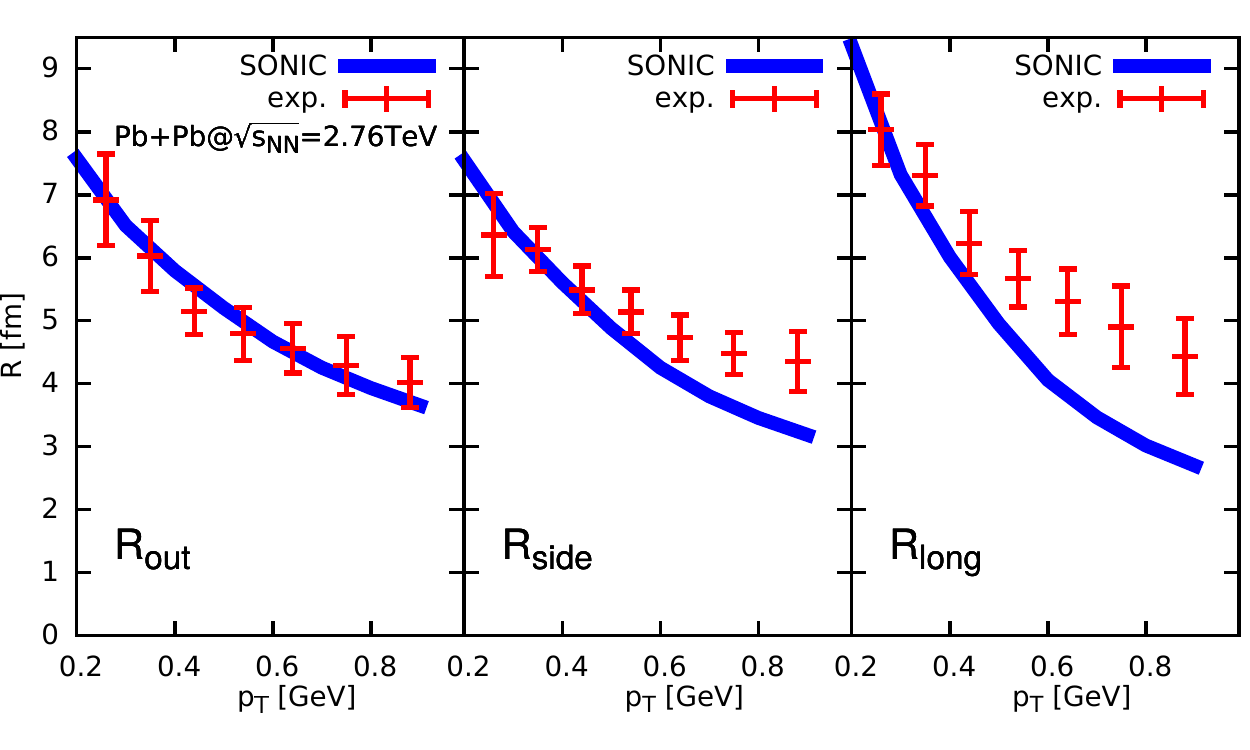}
 \includegraphics[width=0.49\linewidth]{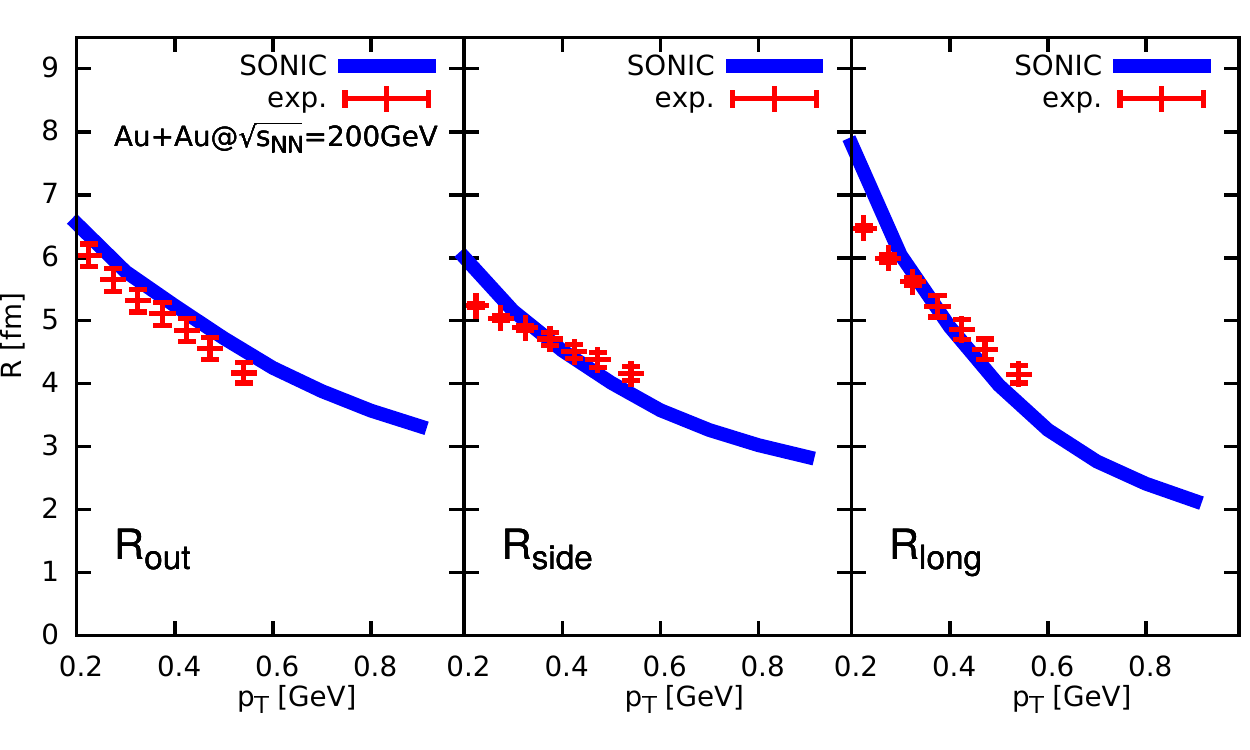}
 \includegraphics[width=0.49\linewidth]{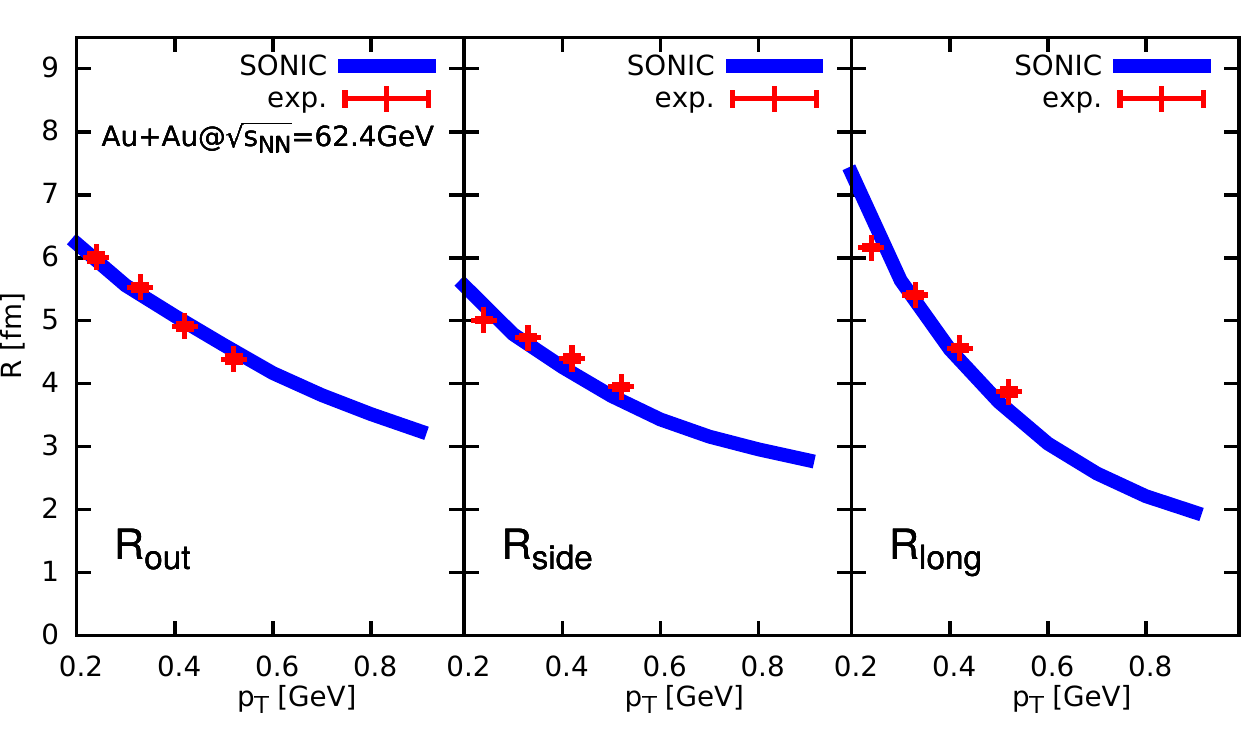}
 \includegraphics[width=0.49\linewidth]{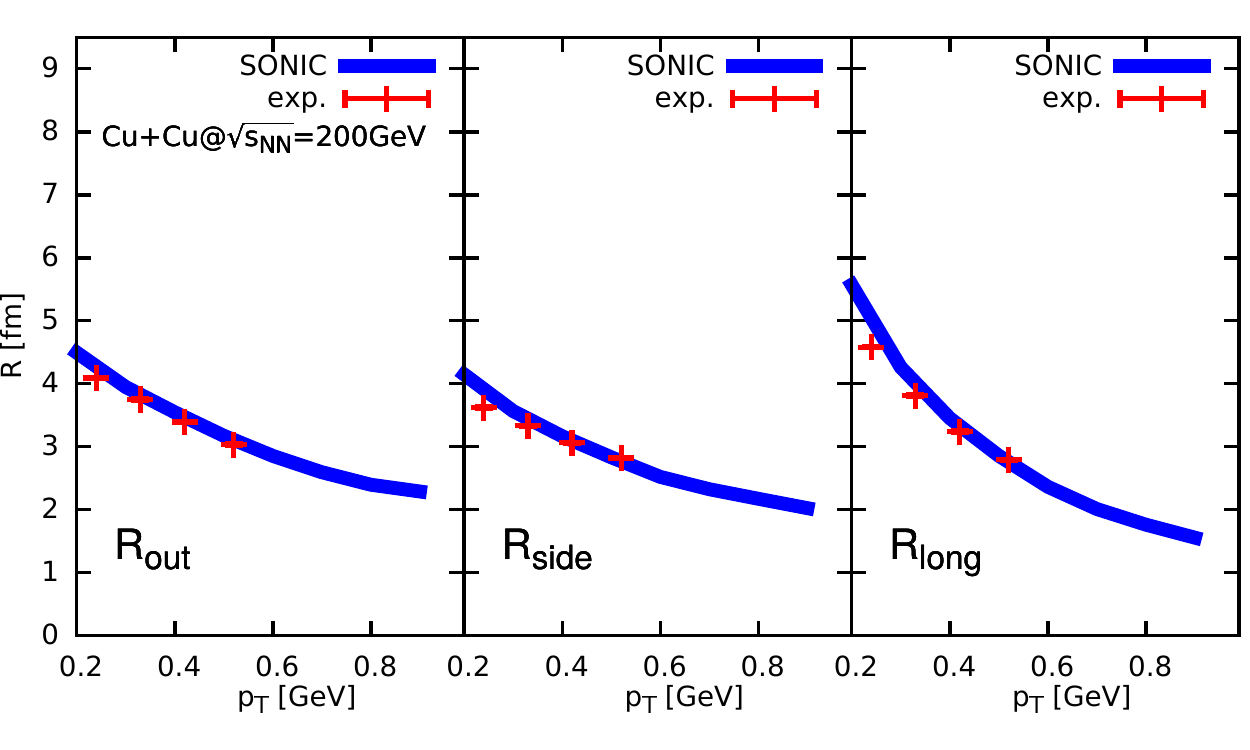}
 \includegraphics[width=0.49\linewidth]{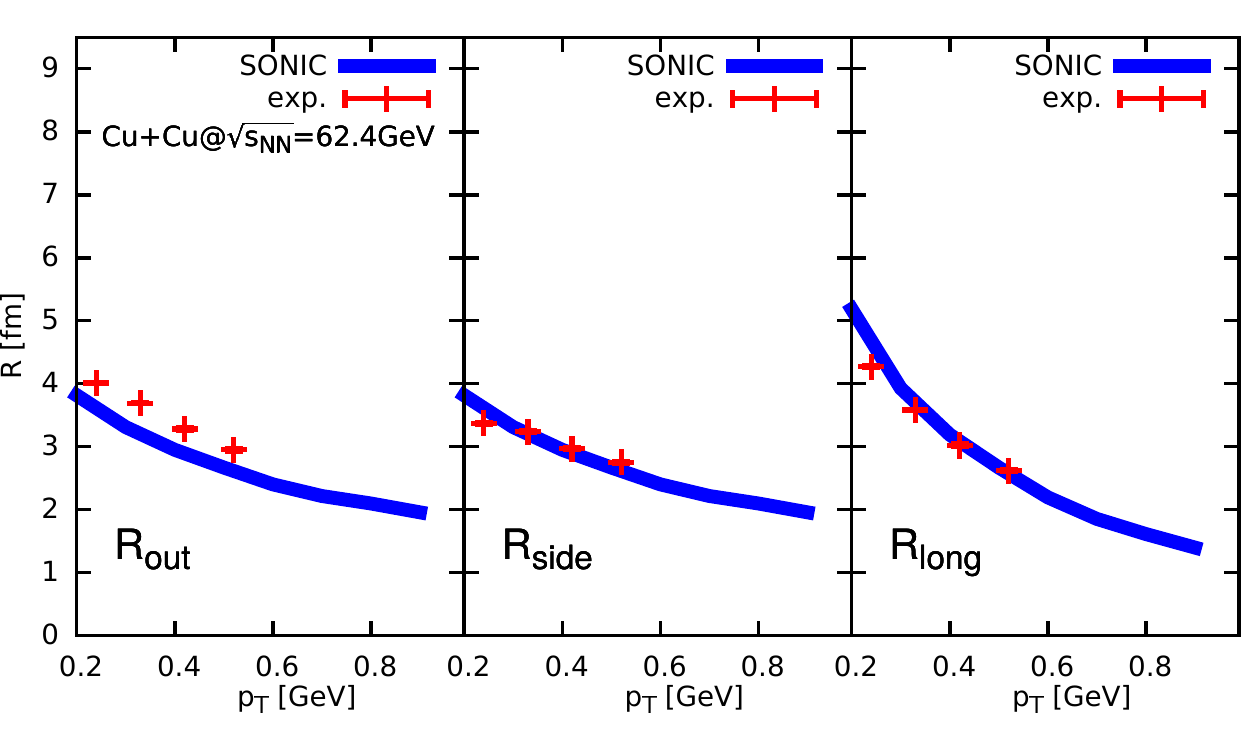}
 \includegraphics[width=0.49\linewidth]{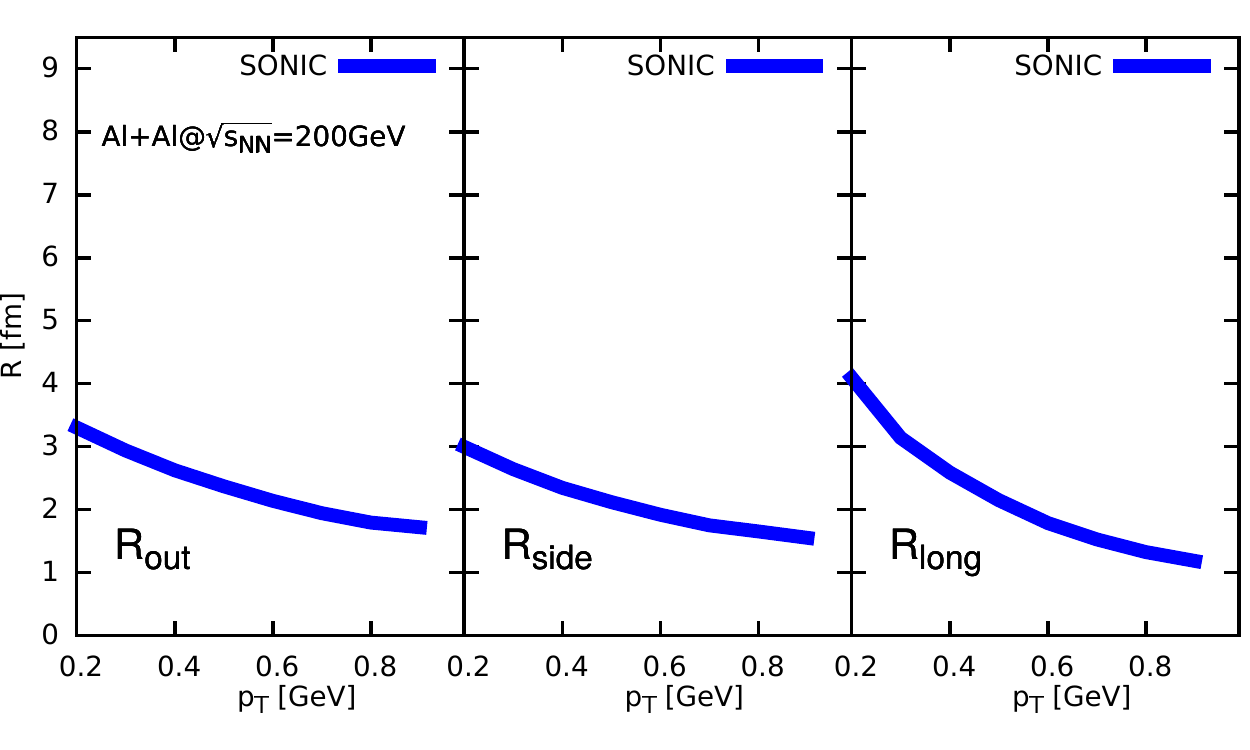}
 \includegraphics[width=0.49\linewidth]{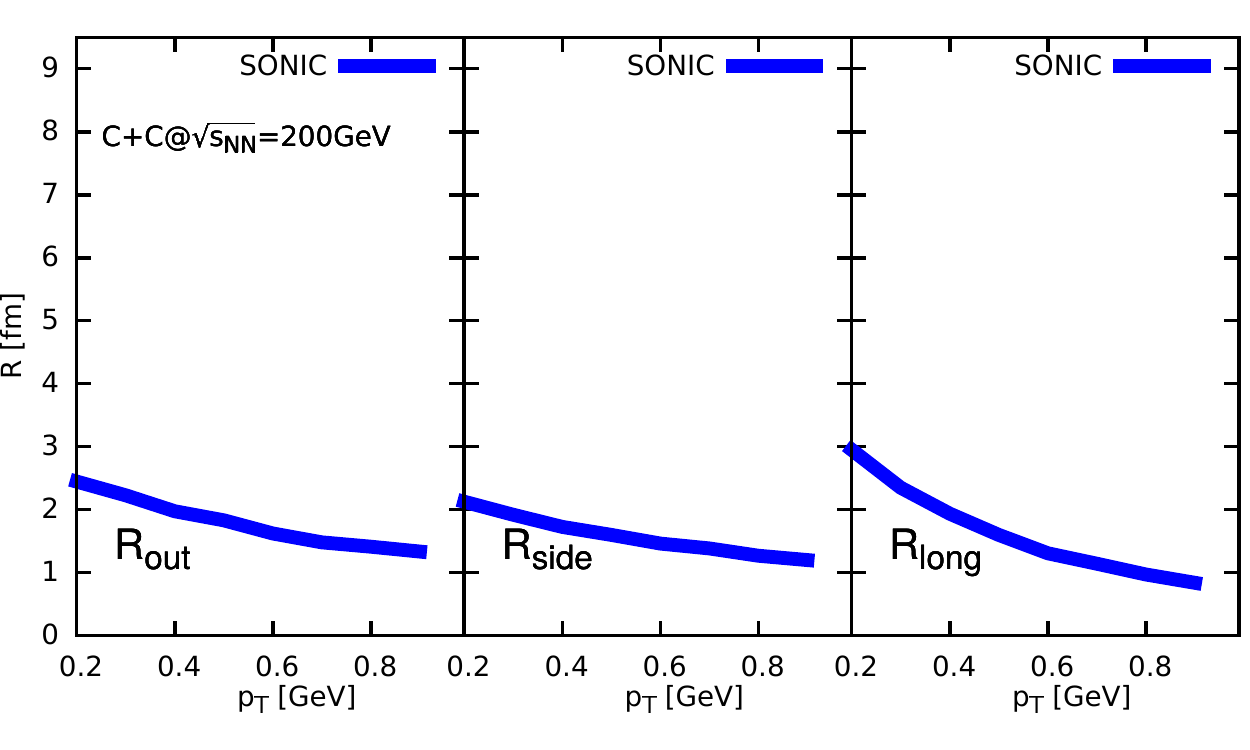}
 \includegraphics[width=0.49\linewidth]{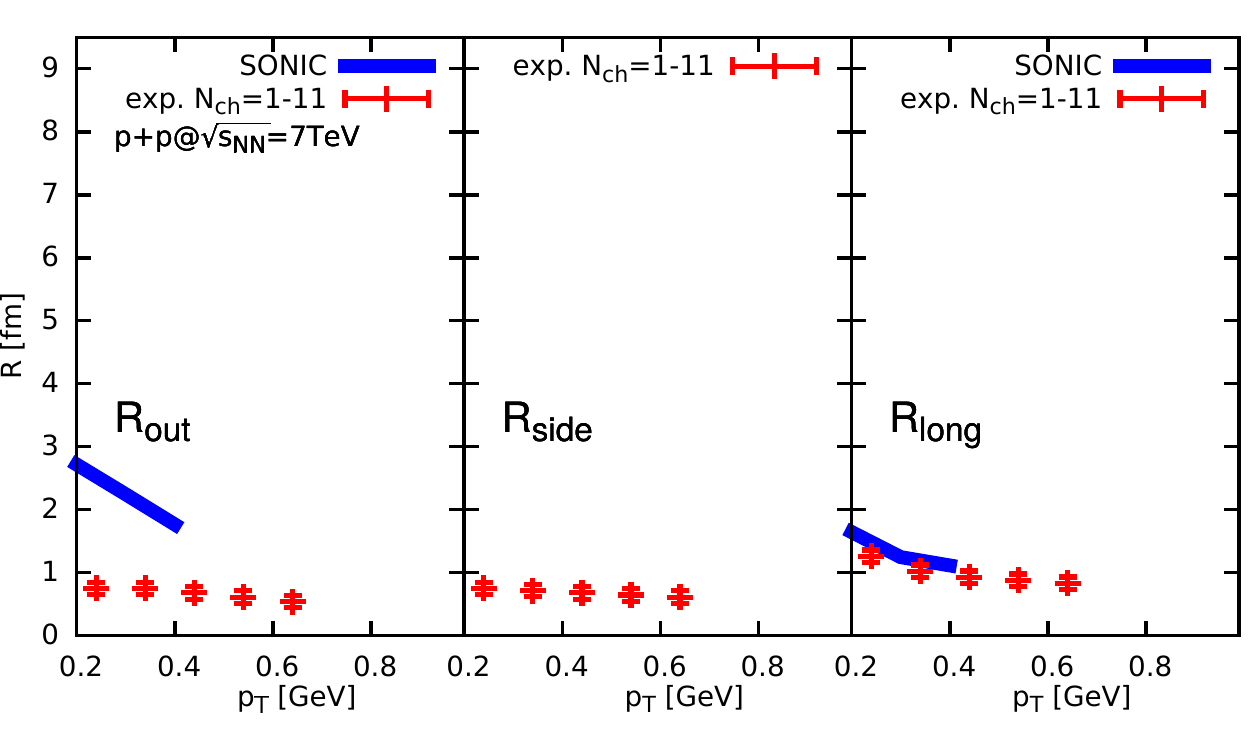}
  \caption{\label{fig:hbt} 
Pion HBT radii for the different collision systems. Shown are model results (\name) and experimental results where available
\cite{Aamodt:2011mr,Adams:2004yc,Abelev:2009tp,Aamodt:2011kd}. For \pp collisions, our numerical method to calculate HBT radii is breaking down, so we only report partial results. Experimental data is for 0-5\% most central events for \pbpb and \auau collisions, 0-10\% most central events for \cucu collisions and minimum-bias events for \pp collisions.
}\end{figure}

Our results for pion HBT radii are calculated as described in \cite{Novak:2013bqa} and the results are shown in Fig.~\ref{fig:hbt} for the different collision systems. Despite some remaining discrepancies between our model results and experimental data, the overall agreement between \name and experiment for different collision energies and systems is striking, given that the inability of standard hydrodynamics to describe the data has been labeled the 'HBT puzzle' in the literature. As noted in Ref.~\cite{Pratt:2008qv}, it is possible to resolve this 'puzzle' by a combination of different ingredients, notably pre-equilibrium flow, viscosity, and a QCD-like equation of state. Since all of these ingredients are naturally incorporated in \name, it is gratifying to observe that the HBT puzzle is no longer a puzzle but rather a (small) discrepancy in some of the data-model comparison.

\begin{figure}[t]
 \includegraphics[width=0.7\linewidth]{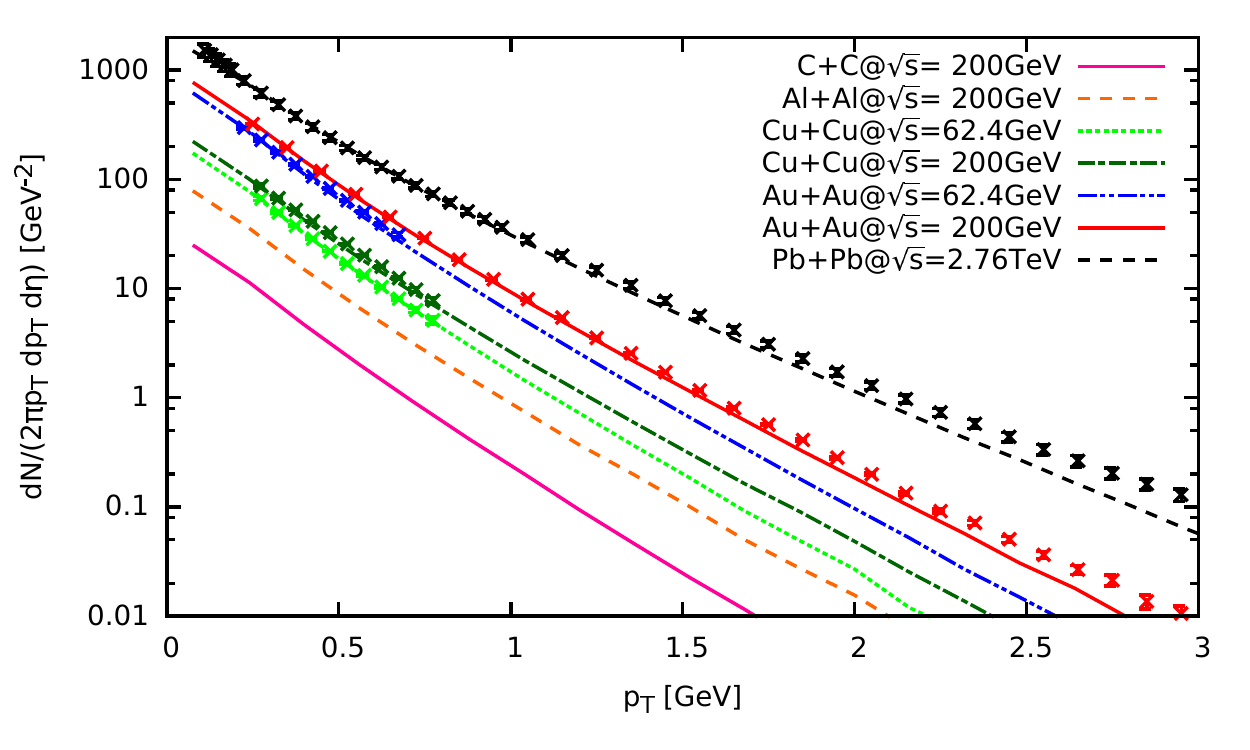}
  \caption{\label{fig:spectra} 
Pion spectra from \name compared to experimental data where available \cite{Adler:2003cb,Abelev:2008ab,Aggarwal:2010pj,Abelev:2012wca}. Experimental data is for 0-5\% most central events for \pbpb and \auau collisions and 0-10\% most central events for \cucu collisions.
}\end{figure}

In Figure~\ref{fig:spectra}, we show the pion transverse momentum spectra for the different collision systems. As remarked above, we do find that with constant values of $\eta/s=0.08$, $\zeta/s=0.01$, and a QCD equation of state, \name provides a good overall description of the available experimental data. Note that the discrepancy in the pion spectra for \pbpb collisions at $p_T>1.5$ GeV was not observed in Ref.~\cite{vanderSchee:2013pia}. The reason is that in Ref.~\cite{vanderSchee:2013pia}, the actual calculation erroneously used a model parameter value of $R=6.48$ fm instead of $R=6.62$ fm (cf. Tab.~\ref{tab:one}) for Pb. Once correcting for this error, we do find slightly less transverse flow in \pbpb collsions, leading to the discrepancy observed in Fig.~\ref{fig:spectra}. However, it is expected that implementing more realistic granular initial conditions will lead to higher transverse flow velocities. This could help to improve the description of experimental data at $p_T>1.5$ GeV in \name in the future.

\section{Conclusions}

We have presented \name, a new super hybrid model for heavy-ion collisions that combines pre-equilibrium flow, viscous hydrodynamics, and hadronic cascade dynamics into one package. \name was used to simulate boost-invariant, central, symmetric collisions of smooth nuclei (Pb, Au, Cu, Al, C) at energies ranging from $\sqrt{s}=62.4$ GeV to $\sqrt{s}=2.76$ TeV. We found that for a QCD equation of state and a choice of QCD viscosity over entropy ratios of $\eta/s=0.08$, $\zeta/s=0.01$, the particle spectra and pion HBT radii were in reasonable agreement with available experimental data. We also made predictions for pion mean transverse momentum and HBT radii for \cc and \alal collisions at $\sqrt{s}=200$ GeV. The 2+1 dimensional space--time evolutions of the  temperature  obtained with \name are publicly available \cite{data} in order to be of use in future studies of jet energy loss or photon emission.

\begin{acknowledgments}
 
This work was supported by the Department of Energy, awards No. DE-SC0008027, DE-SC0008132 and DE-FG02-00ER41244. We would like to thank the JET Collaboration, Chun Shen and Ulrike Romatschke for discussions.

\end{acknowledgments}

\bibliographystyle{apsrev} \bibliography{sphenix}

\end{document}